# Probing the Order Parameter of Superconducting LiFeAs using Pb/LiFeAs and Au/LiFeAs Point-Contact Spectroscopy

*Xiaohang Zhang,[1][*] Bumsung Lee,[2] Seunghyun Khim,[2] Kee Hoon Kim,[2] Richard L. Greene,[1] and Ichiro Takeuchi[3]*

[1]CNAM and Department of Physics, University of Maryland, College Park, Maryland 20742, USA

[2]CeNSCMR, Department of Physics and Astronomy, Seoul National University, Seoul 151-747, South Korea

[3]Department of Materials Science and Engineering, University of Maryland, College Park, Maryland 20742, USA

(Dated on Feb. 20, 2012)

**ABSTRACT**. We have fabricated *c*-axis point contact junctions between high-quality LiFeAs single crystals and Pb or Au tips in order to study the nature of the superconducting order parameter of LiFeAs, one of the few stoichiometric iron-based superconductors. The observation of the Josephson current in *c*-axis junctions with a conventional *s*-wave superconductor as the counterelectrode indicates that the pairing symmetry in LiFeAs is not pure *d*-wave or pure spin-triplet *p*-wave. A superconducting gap is clearly observed in point contact Andreev reflection measurements performed on both Pb/LiFeAs and Au/LiFeAs junctions. The conductance spectra can be well described by the Blonder-Tinkham-Klapwijk model with a lifetime broadening term, resulting in a gap value of ≈ 1.6 meV ($2\Delta/k_BT_C \approx 2.2$).

[*] Email address: xhzhang@umd.edu

**Introduction**

Since the discovery of superconductivity in iron pnictide compounds [1], the symmetry of the superconducting order parameter in these materials [2,3] has been heavily investigated with the ultimate goal of determining the pairing mechanism. To date, there is no consensus on the pairing symmetry which can universally explain the diverse range of physical properties reported in the literature for the entire family of iron-based superconductors. In fact, one emerging picture is that perhaps there are different symmetries for different classes of pnictides with different structures. The multi-band nature of the iron-based superconductors is one possible reason which leads to these differences. The superconductivity has been suggested to arise from a magnetic coupling between electron and hole pockets at the Fermi surface. Such a pairing interaction mediated by magnetic excitations (spin waves) can lead to a multi-band superconductivity with a sign reversal between the electron pocket and the hole pocket, represented by the $s\pm$ wave symmetry [4]. It has also been suggested that orbital fluctuations may play an important role in some iron pnictide compounds, possibly leading to a multiband superconductivity without $\pi$-phase shift between different bands, resulting in the $s++$ wave symmetry [5]. However, the observation of nodal gap structures has been recently reported in several systems such as LaFePO [6], $KFe_2As_2$ [7], and $BaFe_2As_{2-x}P_x$ [8]. These results point to the possible existence of $d$-wave or even $p$-wave pairing symmetry in these materials. But nodes can also be explained within an $s$-wave symmetry [9] where disorder introduced upon doping or substitution in '1111' and '122' iron pnictide superconductors can cause pair-breaking scattering, leading to an "accidental" point or line nodes in the superconducting order parameter.

LiFeAs [10] is a stoichiometric compound in the iron pnictide family with a relatively high transition temperature of $\geq 17$ K and a large residual resistivity ratio up to 50. The spin-density wave (SDW) ordering and the structural transition seen in the parent compounds of the '1111' and the '122' systems are not present in LiFeAs. These distinct features make LiFeAs a unique system for studying the intrinsic electronic properties as well as its pairing symmetry. Despite the absence of a static magnetic transition in LiFeAs, several nuclear magnetic resonance (NMR) studies [11,12] have suggested that impurities or defects in the system may cause antiferromagnetic fluctuations, and pairing through them could still lead to the $s\pm$-wave symmetry. However, angle resolved photoemission spectroscopy (ARPES) measurements [13] have indicated the presence of strong electron-phonon interactions in LiFeAs implying that the superconductivity might be phonon mediated. Moreover, a number of studies including the penetration depth [14], ARPES [15], specific heat [16] and directional thermal transport [17] measurements indicate that the gap structure in LiFeAs is highly isotropic without nodes. Recently, it was reported that the Knight Shift of LiFeAs shows no change at the superconducting transition when magnetic field is applied perpendicular to the $c$-axis [18]. Because this behavior has been previously observed in the unconventional superconductor $Sr_2RuO_4$ [19], such a result brings the spin-triplet $p$-wave symmetry into the discussion as a possible pairing symmetry for LiFeAs [20].

The amplitude of superconducting order parameter is an important measure of the strength of superconductivity. Early small angle neutron scattering (SANS) and ARPES measurements [15] by Inosov *et al*. on LiFeAs have indicated an isotropic gap of ~ 3.1 meV ($2\Delta/k_BT_C \approx 4.0$), suggesting that it is a weakly coupled superconductor. This single gap scenario is also supported by recent thermal transport measurements [17]. A yet another ARPES study [21] by Borisenko *et al*. reported a second gap with a value of 1.0 meV at the hole band, suggesting a double-gap structure for LiFeAs. Although the reported gap size varies, measurements by several other techniques [14,16,22-27] also indicate a double-gap structure for LiFeAs.

In this work, we report on *c*-axis junctions fabricated on large LiFeAs single crystals. Specifically, point contact junctions between an *s*-wave superconductor and a LiFeAs single crystal are studied as a test for the possible presence of the proposed *p*-wave pairing symmetry, and point-contact Andreev reflection (PCAR) spectroscopy measurements are performed to directly probe the size of the superconducting gap of LiFeAs. Previously, we have used the same approach to observe Josephson coupling in *c*-axis Pb/(Ba,K)Fe$_2$Se$_2$ single crystals [28] and determine the superconducting gap in a series of 122 family compounds [29]. The present observation of the Josephson effect in *c*-axis Pb/LiFeAs junctions suggests that the pairing symmetry in this pnictide superconductor is inconsistent with a pure spin-triplet *p*-wave or a pure *d*-wave. Moreover, a superconducting gap is clearly observed in PCAR spectroscopy with both Pb and Au as the counterelectrodes. Fitting the conductance spectra with a modified Blonder-Tinkham-Klapwijk (BTK) model [30,31], the gap value is determined to be ~ 1.6 meV, resulting in a 2$\Delta/k_BT_C$ ratio of ~ 2.2.

**Experimental Results and Discussion**

There have already been numerous experimental and theoretical work on Josephson junctions involving pnictides [32] with implications for the pairing symmetries in this class of superconductors [28,33]. In the present study, the LiFeAs single crystals used were from two batches grown by the Sn-flux method. Fig. 1a shows a typical temperature dependence of the resistivity of the crystals. The resistivity completely drops to zero at about 17 K which is taken as the T$_C$ of the LiFeAs crystals. The sharp transition width (< 1 K) and the high residual resistivity ratio (> 20) attest to the high quality of the single crystals. Details of the growth method and the characteristics of the single crystals are described in Ref. 34. Scanning electron microscopy (SEM) indicates that most single crystals show large terrace-free *ab*-plane surfaces up to several (mm)$^2$ in area. A representative SEM image is shown in Fig. 1b. Our *c*-axis point contact junctions were fabricated on these flat surfaces. The single crystals were mounted and pre-aligned in ambient within about ten minutes so that surface oxidization is minimized. To avoid damaging the surface and to ensure the directional control in junction transport, soft materials, i.e. Pb and Au, were used as counterelectrodes in this study. After measurements, the typical flattened tip area was found to be around $100 \times 100$ $(\mu m)^2$, while no damages to the crystal surface could be detected under an optical microscope.

At 4.2 K, Josephson currents were consistently observed in all *c*-axis Pb/LiFeAs junctions. Fig. 2a shows a typical *I-V* curve of the junctions, which has the resistively-shunted-junction (RSJ) like characteristic with no hysteresis. Under irradiation of a 4.2 GHz microwave field, sharp Shapiro steps were observed in the *I-V* characteristics at voltages corresponding to multiples of *hf*/2*e* as seen in Fig. 2a. The $I_C R_N$ products of the junctions at 4.2 K ranged from ≈ 10 μV to ≤ 1 mV for the eight junctions we studied. By applying a magnetic field, the observed critical currents are found to be completely suppressible showing Fraunhofer-like patterns with a typical modulation period of about 7 Oe. Using an estimated penetration depth of 200 nm for LiFeAs and the penetration depth of 50 nm for Pb at 4.2 K, this modulation period corresponds to an effective junction width (W) of ~ 10 μm at the interface of the flattened Pb tip and the LiFeAs *ab*-plane surface [35]. The $R_N$ values of our junctions range from 10 mΩ to 150 mΩ. If we assume that the current flows through the entire nominal contact region [i.e. the flattened tip area of ~ $100 \times 100$ $(\mu m)^2$], the interface resistivity would be ≥ $10^{-6}$ Ωcm$^2$, which is generally too large to carry Josephson currents. However, if we use the effective junction area of ~ $10 \times 10$ $(\mu m)^2$, the interface resistivity drops to ~ $10^{-7}$-$10^{-8}$ Ωcm$^2$, a range where Josephson coupling is more likely to take place [36].

The effective junction dimensions of ~ $10 \times 10$ $(\mu m)^2$ and the observed Josephson current range give the Josephson penetration depth ($\lambda_J$) in the range of 10 - 25 μm, placing our junctions in the intermediate ($W/\lambda_J \sim 1$) to the small junction limit ($W/\lambda_J < 1$) [28], which in turn, is consistent with the shape of the observed magnetic diffraction patterns (not shown). The actual effective area being in the range of 100 (μm)$^2$ rather than the entire flattened tip area is also consistent with the fact that the very surface of LiFeAs is sensitive to degradation and that only a fraction of the large terrace-free surface is actually "active" (and non-insulating) to be probed by point contact transport.

The observation of the Josephson coupling in *c*-axis junctions between the *s*-wave Pb tip and LiFeAs single crystal implies the predominance of one sign in the order parameter in this pnictide superconductor, which suggests the existence of an s-wave component in the order parameter. Therefore, the pairing symmetry in LiFeAs is not pure *d*-wave or pure spin-triplet *p*-wave. This is a main conclusion of the present work.

The PCAR spectroscopy is a versatile technique which allows direct probing of the gap structure in superconductors. Since the discovery of superconductivity in iron pnictides, PCAR has been applied to a variety of compounds in the '1111' and the '122' families [29,37-47]. Both single-gap and double-gap features have been reported. Compared to other experimental techniques [3] in which the multiple-gap superconductivity is evident, the PCAR spectroscopy usually shows less clear features for large gaps [39-44,46-48], making the precise determination of the gap value non-trivial. On the other hand, it is also well known that impurities, local stress, nonuniformity of contacts, etc. can potentially induce artificial features on the conductance spectrum [49]. Specifically, since both the '1111' and the '122' pnictide superconductors are doped systems, the scattering at the interface in point-contact studies can complicate the transport. Compared to such doped pnictide superconductors, stoichiometric LiFeAs single crystals exhibit high residual resistivity ratios and low densities of impurities or defects suggesting that PCAR measurements performed on this superconductor might yield spectra with clearer gap features.

The observation of the RSJ-like Josephson coupling in Pb/LiFeAs junctions implies that the contact is highly transparent. Such transparent contacts translate to a relatively small barrier strength Z, as discussed below, and are the key to a precise fit to the conductance spectra and for determining the gap values in PCAR studies. Our microscopic picture of the contact is that the current is flowing through multiple nanoscale ballistic contacts within the 10 μm×10 μm area. In PCAR studies, such multiple-path contacts are frequently proposed based on the fact that conductance spectra can be well described by the ballistic Blonder-Tinkham-Klapwijk (BTK) model. Fig. 2b shows a differential resistance spectrum measured on a Pb/LiFeAs junction at 4.2 K. At zero bias, the junction resistance goes to zero, consistent with the observed Josephson current. The entire curve was normalized by a fit to the smooth high bias voltage ($|V| > 5$ mV) spectrum, resulting in a normalized conductance spectrum as shown in Fig. 2c. The normalized spectrum shows a conductance enhancement of about 25%, suggesting again that the contact is relatively transparent. No obvious additional features corresponding to multiple Andreev reflections [50] were observed on the spectrum. Compared to the spectrum obtained at 8 K, the conductance enhancement occurs at a nearly identical bias voltage, indicating that this feature indeed has its origin in the superconductivity of LiFeAs.

To quantitatively describe the obtained conductance spectrum, we use the single-gap BTK model with a lifetime broadening term. In this model, a dimensionless parameter Z is introduced to describe the barrier strength while an energy term, Γ, is used to represent the lifetime broadening due to inelastic scattering [31]. The best fit to the spectrum shown in Fig. 2c suggests that the size of the superconducting gap in LiFeAs is ≈ 1.65 meV.

Fig. 3a shows a set of resistance spectra obtained on a Au/LiFeAs junction at various temperatures. With the normal metallic Au tip, potential complexity due to multiple Andreev reflections is eliminated. A superconducting gap feature is clearly evident in the resistance spectra, and the feature gradually vanishes when the temperature is raised to approach the transition temperature of LiFeAs. The high-bias spectra show negligible change with increasing temperature, which implies that the normalization obtained by a fit to the high-bias spectra at low temperatures is equivalent to the normalization by the normal state spectrum (i.e. $T > T_C$). The normalized conductance spectra are shown in Fig. 3b. A dip at zero bias is observed in the low temperature conductance spectra, which usually indicates that the junction is in an intermediate regime between tunneling and metallic contact, represented by a finite Z value in the BTK model. By applying the modified BTK model, single-gap fits to the data are obtained (Fig. 3b). In particular, the results indicate a gap value of 1.6 meV at 4.2 K, consistent with the value obtained on Pb/LiFeAs junctions (Fig. 2c). With these values, we arrive at the $2\Delta/k_B T_C$ ratio of about 2.2 for LiFeAs, consistent with the values obtained from penetration depth (Ref. 17 and 23), nuclear quadrupole-resonance (NQR) and NMR measurements (Ref. 22), but lower than the value for weakly-coupled BCS superconductors (3.5) or the single gap value obtained in Ref. 15. The temperature dependence of the extracted gap value plotted in Fig. 4 is in good agreement with that of BCS *s*-wave superconductors.

The conductance spectra in this study can be well described by the single-gap BTK model with no additional features that have been frequently reported in previous measurements on the 122 pnictide superconductors and usually attributed to multiple-gap structures. However, it is difficult to definitively rule out the double-gap scenario from our measurements. If, for instance, there are two gaps close to each other in magnitude, the detailed feature of the second gap could be smeared out by thermal fluctuation and inelastic scattering, leading to a large uncertainty in determining the second gap value. We have indeed attempted rigorous double-gap fits to our spectra which consistently gave us the results that the weighting factor for one of the gaps is close to zero. Furthermore, we note that regardless of the presence or the value of the second gap, a superconductor gap with a value of about 2.0 meV is consistently observed from the conductance spectra obtained in this study.

Previous measurements by SANS and ARPES [15] as well as directional thermal transport [17] have indicated an isotropic gap for LiFeAs, and a high field Hall effect study in the normal state has shown a linear field dependence up to 10 tesla with a negative Hall coefficient [51]. Considering that LiFeAs is a stoichiometric system with a very low impurity level, these results together may be suggestive of the fact that electrons are the predominant charge carrier in the normal state due to a low Fermi velocity ($v_F$) of the holes. In fact, a van Hove singularity and a flattened or narrowed hole band at the Fermi level has been revealed in a recent ARPES study [21]. The results suggest that the hole band has a negligible effect on the superfluid density due to the small $v_F$. If we adopt this picture, it is not surprising that the opening of a superconducting gap at hole pockets could be detected by ARPES and some other techniques, but not by PCAR spectroscopy measurements because of the negligible contribution of the hole band in the superfluid density. Accordingly, the possible gap features that arise from the hole pockets are expected to be small in PCAR spectroscopy measurements. Based on this, the determined gap here is likely from the electron pockets. Moreover, a much larger gap value was obtained in a recent $H_{C2}$ study [52] on crystals from the same batch of samples as the present work. The two distinct gap values obtained in Ref. 52 and the present work may be indicative of the fact that LiFeAs is a double-gap superconductor with the larger order parameter becoming more evident when the smaller one is suppressed by high magnetic fields. These results are consistent with the van Hove singularity picture proposed in Ref. 21.

**Summary**

We have performed a systematic study on the superconducting order parameter of LiFeAs single crystals through *c*-axis point contact junctions with Pb and Au tips. The clear Josephson current observed in the *c*-axis junctions between an s-wave superconductor and a LiFeAs single crystal strongly suggests the existence of an s-wave symmetry in this pnictide superconductor. A superconducting gap is clearly identified in point contact Andreev reflection spectroscopy studies. By applying the modified BTK model, a gap value of about 1.6 meV is determined, giving the $2\Delta/k_BT_C$ ratio of about 2.2. Taken together with previous reports on this compound, the present results suggest that the identified gap is from the electron pocket.

**Acknowledgement**

The authors acknowledge fruitful discussions with J. Paglione, K. Jin, N. P. Butch and S. R. Saha. Work at UMD was partially supported by the NSF under Grant No. DMR-1104256 and by the AFOSR-MURI under Grant No. FA9550-09-1-0603; I.T. is also supported by NSF MRSEC at UMD (Grant No. DMR-0520471); work at SNU was supported by national creative research initiatives (2010-0018300) and the Fundamental R&D Program for Core Technology of Materials by MKE.

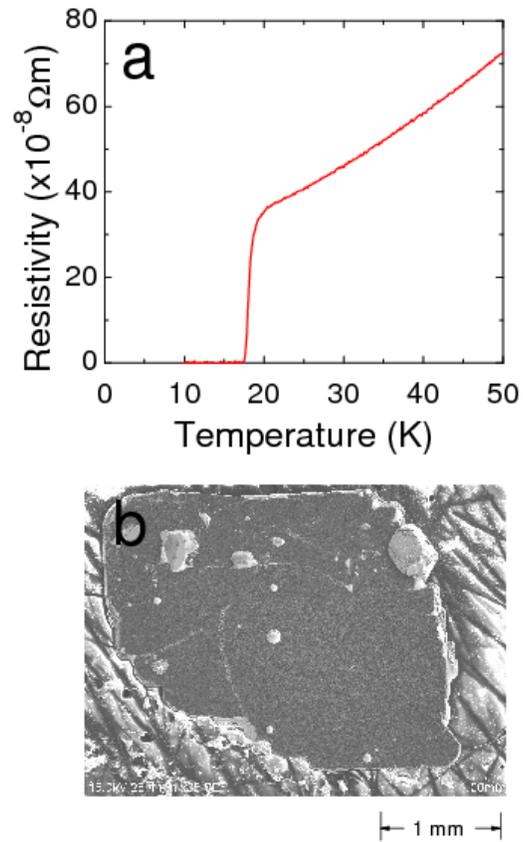

FIG. 1. (Color online) (a) Temperature dependence of the *ab*-plane resistivity of a LiFeAs single crystal. The transition temperature ($T_C$) is 17 K; (b) A representative SEM image of a single crystal (mounted on conducting carbon tape) shows presence of large flat areas on the surface of the crystal.

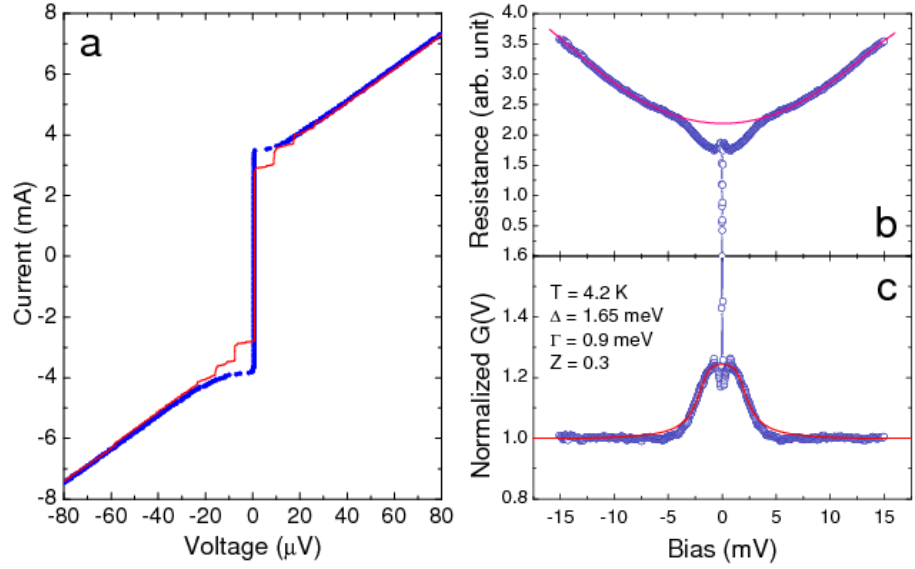

FIG. 2. (Color online) (a) A typical RSJ-like Josephson I-V characteristic (blue dots) obtained on a Pb/LiFeAs junction at 4.2 K. Upon irradiation of a 4.2 GHz microwave field, a series of Shapiro steps are clearly observed (red line); (b) Junction resistance (d$V$/d$I$) measured by lock-in detection in a large bias voltage range. At zero bias, the junction resistance drops to zero due to the Josephson current. The red line is a fit to the high-bias data which is used for normalization; (c) the normalized conductance (dI/dV) spectrum from (b). The red line is a single-gap BTK fit with a life time broadening term $\Gamma$. The fitting parameters are listed in the figure.

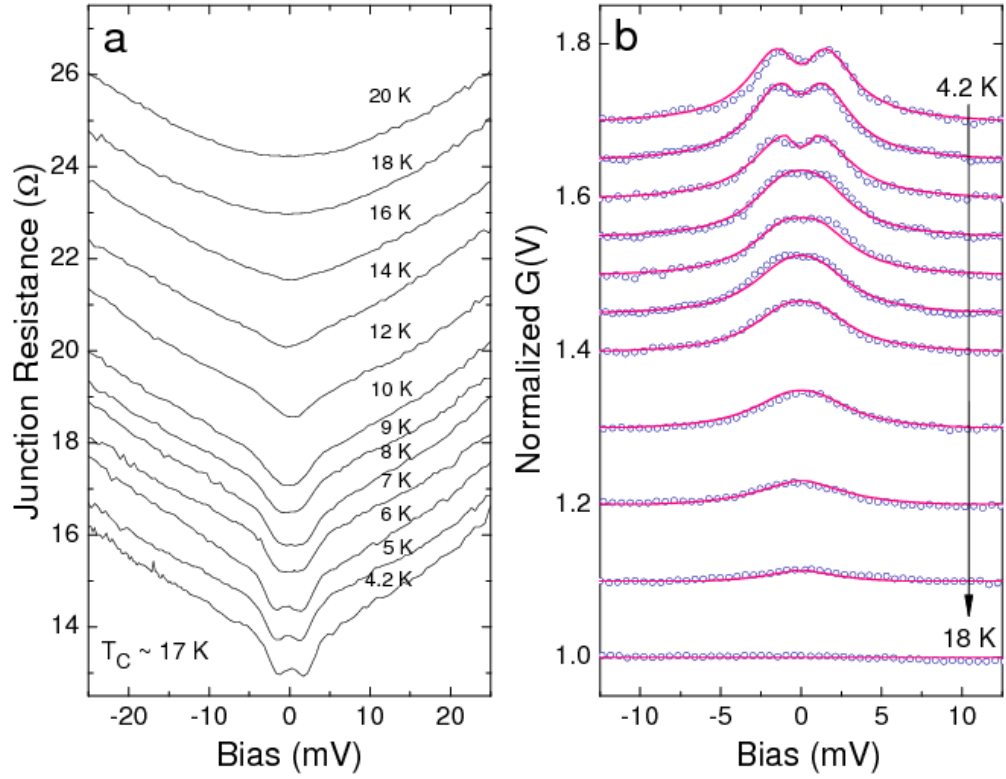

FIG. 3. (Color online) (a) Raw resistance spectra obtained on a Au/LiFeAs junction at different temperatures. The curves are vertically shifted for clarity except the one at 4.2 K. (b) Normalized conductance spectra with BTK fits (lines). The curves are shifted for clarity except the one at 18 K. The obtained gap value at 4.2 K is 1.6 meV.

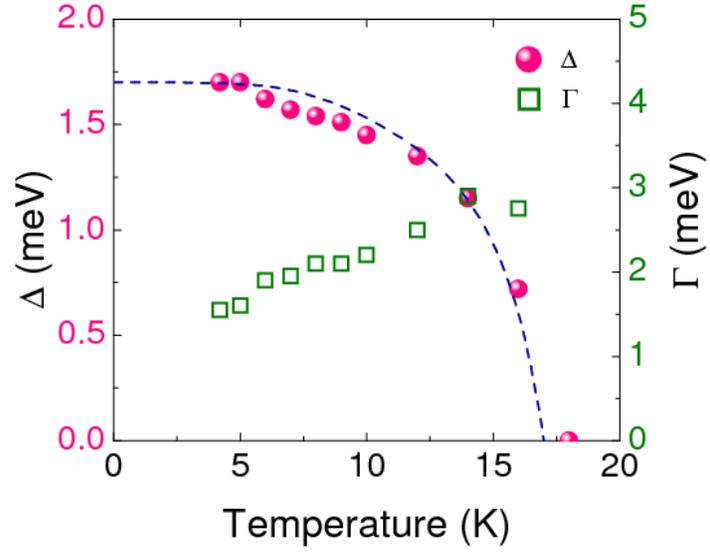

FIG. 4. (Color online) Temperature dependence of the gap value (dots) and the lifetime broadening term (squares) obtained by modified BTK fits to data taken from a Au/LiFeAs junction (shown in Fig. 3). The dashed line is a fit to the temperature dependence of the BCS superconducting gap.